\documentclass{article}
\usepackage{spconf,amsmath,graphicx}
\usepackage{multirow}
\usepackage{booktabs}
\usepackage{float}
\usepackage[normalem]{ulem}
\useunder{\uline}{\ul}{}
\usepackage[caption=false]{subfig}

\title{Learnt Deep Hyperparameter selection in Adversarial Training for compressed video enhancement with perceptual critic}
\name{Darren Ramsook$^{\star}$ \qquad Anil Kokaram$^{\dagger}$ }
\address{SigMedia Group, \\ Trinity College Dublin \\ 
\textit{$^{\star}$ramsookd@tcd.ie}, \textit{$^{\dagger}$anil.kokaram@tcd.ie}}
\begin{document}
\ninept
\maketitle
\begin{abstract}
Image based Deep Feature Quality Metrics (DFQMs) have been shown to better correlate with subjective perceptual scores over traditional metrics. The fundamental focus of these DFQMs is to exploit internal representations from a large scale classification network as the metric feature space. Previously, no attention has been given to the problem of identifying which layers are most perceptually relevant. In this paper we present a new method for selecting perceptually relevant layers from such a network, based on a neuroscience interpretation of layer behaviour. The selected layers are treated as a hyperparameter to the critic network in a W-GAN. The critic uses the output from these layers in the preliminary stages to extract perceptual information. A video enhancement network is trained adversarially with this critic. Our results show that the introduction of these selected features into the critic yields up to 10\% (FID) and 15\% (KID) performance increase against other critic networks that do not exploit the idea of optimised feature selection. 
\end{abstract}
\section{Introduction}
\label{sec:intro}
Lossy compression is necessary to support the current infrastructure of modern media applications. However it introduces a wide range of spatial and temporal artifacts that negatively impact the end user quality of experience~\cite{9035388}. Statistical techniques for artifact removal as a post-processing step has been a long studied research topic~\cite{SHEN19982}.  More recently, Deep Neural Networks (DNNs) have been used in different areas of the compression pipeline, from end-to-end learnt compression~\cite{toderici2017full}, to purely post-processing~\cite{zhang2021video} and mixed post/pre-processing~\cite{guleryuz2021sandwiched}. The use of DNNs for compressed video enhancement has high potential for its ability to exploit perceptually salient metrics in the training process through the use of generative adversarial networks (GANs), something which is difficult to exploit with traditional statistical methods.  

The use of transfer learning for crafting Deep Feature Quality Metrics (DFQMs) has been an emerging area of interest in the Deep Learning Community. Image DFQMs have been shown have a high correlation with human based subjective scores~\cite{ding2020image}. The underlying features used to generate these scores are the internal representations of high performing pre-trained image classification networks e.g. VGG~\cite{simonyan2014very}. LPIPS~\cite{zhang2018unreasonable} has been shown to outperform traditional quality metrics such as $\ell_{2}$, SSIM and FSIM. FloLPIPS~\cite{danier2022flolpips}, an extension of LPIPS that emphasizes areas of motion in a patch, has outperformed other traditional metrics including VMAF and MS-SIM. 

\begin{figure}
    \centering
    \includegraphics[width=\columnwidth]{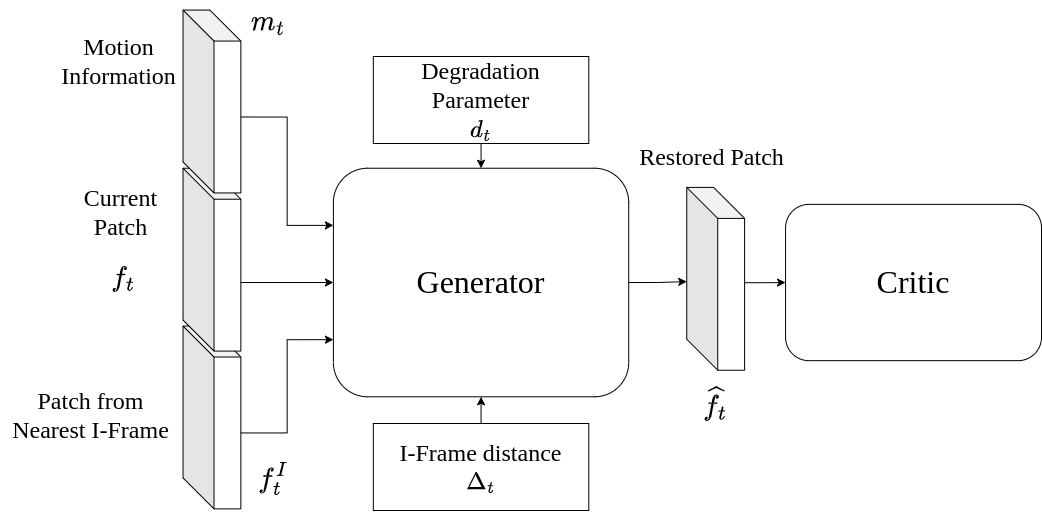}
    \caption{\em Proposed W-GAN setup for compressed video enhancement. The generator take in motion information, the current patch and the patch from the nearest I-frame. The critic network includes selected extracted outputs EfficientNetB3 to assist with restoration.}
    \label{fig:Overall}
    \vspace{-1.8em}
\end{figure}

While these DFQMs are used in wide range of applications for assessing algorithmic performance and also in the loss-functions for training networks, the features that are used to generate these new metrics lack explainability. This is a characteristic hallmark of deep neural networks, however there are some approaches that can be used to help segmenting deep features and their importance in the current application. 
\vspace{-1em}
\subsubsection*{Contributions of this paper}
This paper outlines the use of perceptually relevant deep features in the creation of a neural critic for a Wasserstein-GAN (W-GAN) with gradient penalty~\cite{gulrajani2017improved}. This neural critic is then used for training a compressed video enhancement network using data that is compressed with the H.265 codec. The following contributions are made:
\begin{itemize}
    \item A method for evaluating deep features of a pre-trained image classification network to isolate feature blocks that have perceptually relevant features.
    \item The design of a critic network that uses these selected feature blocks as a primary input and an associated compressed video enhancement network that is trained adversarially with the critic. 
\end{itemize}

\vspace{-1em}
\subsection{Related Work}
Perceptually relevant DNN approaches have been recently used within the compression pipeline. Chen et al.~\cite{https://doi.org/10.48550/arxiv.2007.02711} investigates the use of loss functions that include perceptual metrics (VMAF, SSIM or MS-SIM) alongside MSE for training an end-to-end DNN image codec. Their results show more than 20\% coding gain with VMAF in the loss function when compared to a pure MSE loss. There has been sufficient work that show DFQMs~\cite{ding2020image, zhang2018unreasonable, danier2022flolpips} are more closely related to human subjective scores. This suggests that using the features that these metrics are dependent on has the potential to yield better results beyond using hand crafted metrics like VMAF. In our proposed model that focuses on post-processing for videos, we utilize extracted prior deep features from EfficientNetB3 directly in the neural critic of our W-GAN for training. We do not optimize for any other traditional metrics or DFQMs directly. 

Mohammadi et al.~\cite{mohammadi2022perceptual} presents an investigation of the perceptual impact of changing the optimization metric of the loss function for a DNN image codec. Results indicate that training with the DFQM DISTS~\cite{ding2020image} and MS-SSIM yields highest perceptual gain. Images that were compressed with DISTS were selected with the highest winning frequency of 21\% when compared to a total of 9 different optimization schemes. In~\cite{mohammadi2022perceptual}, the entire set of deep features is used for the DFQM. In our work we instead identify the subset of features which are most relevant and deploy them in a more efficient adversarial training approach. 

Ma et al.~\cite{https://doi.org/10.48550/arxiv.2011.09190} presents a post-processing GAN based approach that uses a mixture of multiple objective metrics. Their loss function incorporates the distance between complete deep features of a degraded-reference pair directly. In our approach, we do not use a difference between deep feature pairs. Instead, we improve performance by only using a subset of features that are perceptually relevant. These features are given to our critic network. This allows the critic to learn distinguishing features between deep representations of clean and degraded patches. 

\vspace{-0.9em}
\section{Deep Feature Selection}

\begin{figure}
    \centering
    \includegraphics[width=0.9\columnwidth]{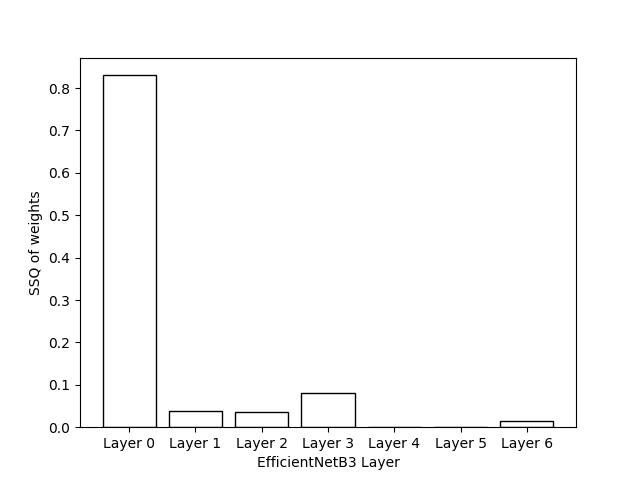}
    \caption{\em Averaged weights per layer in RDM analysis. Based on this measure of importance, layers $0$ and $3$ were selected as the top two most significant internal representations. This follows the notion that layers closer to the input of classification networks encode higher perceptual information.}
    \label{fig:barplot}
    \vspace{-1.8em}
\end{figure}

DFQMs such as LPIPS, FloLPIPS and DISTS are all dependent on the internal representations of pre-trained classification networks. The deep representation of an image is generated by extracting the output of intermediate layers in the network. Features are usually collected at the final output of the convolution filter bank operation of that layer.

These features can be represented by a set of rank 3 arrays whose height and width depends on the size of the input image. The number of channels in the feature array is dependent on the number of convolution filters used at that layer and can be represented by the layer depth. For an arbitrary image input $\tilde{I}$, the set of generated features $F$ can be represented by $F_{z}(\tilde{I}) \in F(\tilde{I})$, where $z$ represents the depth at which a feature is accessed. 

Historic DFQMs ($Q$)~\cite{ding2020image, zhang2018unreasonable, danier2022flolpips} are based on comparing the internal representations of a reference ($x$) and degraded ($y$) patch, $Q(F(x), F(y))$. A key point is that the $F_z$ are extracted from {\em frozen} layers of a network trained for classification. Hence not all of the features will have significant relevance for perceptual criteria. Therefore,
in our proposed method, our critic utilizes only a subset of perceptually relevant features $F^{\alpha}$ from a single input $\hat{x}$: $F^{\alpha}(\hat{x}) \subset F(\hat{x})$. Discovering which subset is perceptually relevant is one of our main contributions. The intuition is that there is a set of compression invariant features that can be used by the W-GAN in restoration of a given patch. In addition, by selecting this subset we reduce the dimensionality of Q() and thus improve computational performance of the critic network by allowing to focus on a subset of key features.

A dataset of 99 reference $512\times512$ patches was used to generate deep feature representations from the classification network EfficientNetB3~\cite{tan2019efficientnet} pre-trained on ImageNet. This network was chosen for its relatively small number of parameters and increased performance over other classificaton networks. The 99 patches were compressed using the H.265 codecs at 6 linearly spaced constant rate factor (CRF) points in the operating range. This resulted in 99 reference and 594 degraded full feature representations. 

Each $F$ for a single patch contains rank-3 arrays that total $\approx$ 3 million elements. Traditional statistical techniques for dimensionality reduction proves intractable for these high element multi-dimensional feature arrays. To overcome this limitation, we take inspiration from neuroscience~\cite{KRIEGESKORTE2013401, SupervisedMay}. In~\cite{SupervisedMay}, inferior temporal (IT) response to images presented to humans were compared to responses from layers in a pre-trained classification network when presented with those same images. Their goal was to discover which CNN layers were able to best model the human response. To do that, the authors exploited a dissimilarity measure \cite{KRIEGESKORTE2013401} for the CNN layer outputs which was then used to predict the same dissimilarity response from the human stimulus. In our case we are interested in discovering the specific layers in our representation (EfficientNetB3) which best enable the critic network to discriminate between degraded and clean video. In a sense the stimulus here is the input images and the response is the output of our CNN layers. Hence we follow the same strategy as the neuroscience case. 

We first define Representational Dissimilarity Matrices (RDMs) for responses $F_z$ from layer $z$ and ${\bf F}$ from all layers, as follows.
\begin{align}
    R^{Y}(i,j) & = d\bigl({\bf F}(y^{i}), {\bf F}(y^{j})\bigr) \\
    R^{{}_{c}X_{z}}(i,j) & = d\bigl(F_{z}(x^{i}_{c}), F_{z}(x^{j}_{c})\bigr)
\end{align}
where $d(\dot, \cdot)$ is the euclidean distance, $y_k$ is the $k$th patch in our clean dataset and $x^k_c$ is the $k$ th patch in the datset degraded through compression with CRF $c$. Hence $F_z(x^i_c)$ is the response of the $z$th layer to the degraded patch $i$. $R^{Y}(i,j)$ is therefore a measurement of the auto-dissimilarity between all the clean patches in the dataset, while $R^{{}_{c}X_{z}}(i,j)$ is the auto-dissimilarity across the degraded patches. 

By building a linear predictor for $R^{Y}(i,j)$ based on linear combination of elements across layers in $R^{{}_{c}X_{z}}(i,j)$, we can select layers based on significance for prediction. The linear predictor is defined through weights ${}_{c}\beta_{z}$ applied to each layer/degradation element in $R^{{}_{c}X_{z}}$  as follows.
\begin{equation}
\hat{R}^{Y}({}_{c}\beta_{z}) = 
     \sum_{\forall z}\sum_{\forall c}{}_{c}\beta_{z}R^{{}_{c}X_{z}}, 
\end{equation}
We estimate ${}_{c}\beta_{z}$ by minimising the cosine distance between $\hat{R}^{Y}({}_{c}\beta_{z})$ and $R^{Y}$. We apply the additional constraint of non-negative weights with the BFGS algorithm\cite{9781118723203}.

By averaging the estimated weights across degradation level $c$ (6 in our case) we estimate the relative importance of each layer. This is shown in Figure~\ref{fig:barplot}. Layers 0 and 3 of EfficientNetB3 are therefore chosen for our critic network as they clearly encode most perceptual information.


\begin{figure}[]
\centering
\subfloat[\em Critic. An input patch is processed by EfficientNetB3 and only layers 0 and 3 (purple) are used in producing output scores. Computation stops at layer 3, other layers are included for clarity.]{%
  \includegraphics[clip,width=\columnwidth]{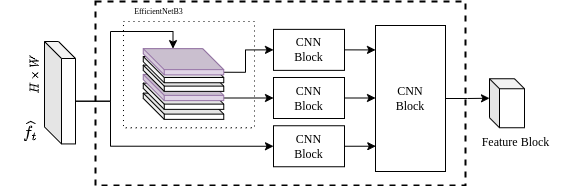}%
}

\subfloat[\em{
Generator. Motion magnitude and direction $m_t$ of the current frame is concatenated with the current RGB patch $f_t$. The corresponding patch from the nearest intra-coded frame is fed into a CNN block (blue). The resulting features are concatenated with the center of the UNet (red). These features are then fed into an attention module (CBAM~\cite{woo2018cbam}). The restoration parameter $d_t$ guides the enhancement network and $\Delta_{t}$ is a weighting factor that scales the features extracted from the I-frame.
}]{%
  \includegraphics[clip,width=\columnwidth]{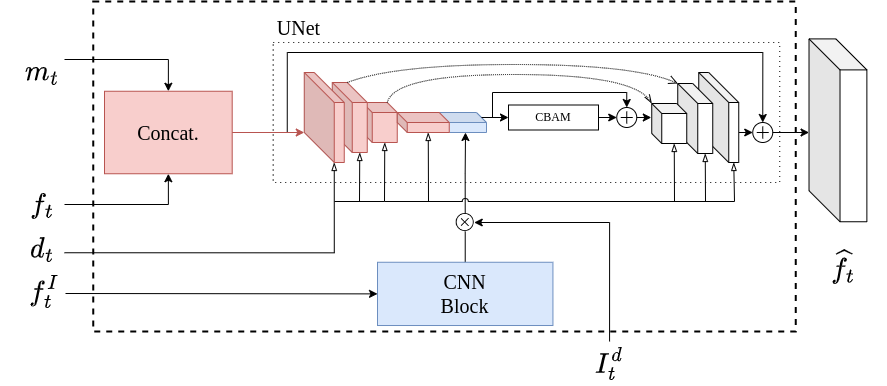}%
}

\caption{{\em Proposed Critic (a) and Generator (b) networks.}}
\label{fig:networks}
\end{figure}

\section{Network Architecture}
 \vspace{-1em}

\textbf{Critic}: Our critic network takes a patch as input and produces a feature block that represents the patch degradation level. This is adapted from the PatchGAN architecture~\cite{isola2017image}. The main contribution of our critic network is the introduction of the selected EfficientNetB3 layers as a primary input. This is shown in Figure~\ref{fig:networks}(a). 

The extracted features are processed in parallel by CNN Blocks. The input frame is also processed directly to capture information that EfficientNetB3 may have missed. These are then concatenated and processed further to produce the feature block.
There is no activation function in the final layer. This follows the requirement of a W-GAN for the critic output to be unbounded.

To compare our proposed critic with optimal layers ($C_{opt}$), we also train with critic networks that:
\begin{enumerate}
    \item Uses the output from layers 5 and 6 of EfficientNetB3 ($C_{5,6}$).
    \item Does not use the EfficientNetB3 module, and hence the entire critic is fully trainable ($C_{t}$).
\end{enumerate}

\noindent\textbf{Generator}: Our generator (Figure~\ref{fig:networks}(b)) uses a UNet structure as the main 
network. Motion vector information (magnitude and direction) are concatenated with the input RGB patch. This is then fed into a UNet with skip connections. Features from the nearest Intra-coded frame (I-frame) are extracted by a CNN block and is concatenated with the center of the UNet. It is assumed that the nearest I-frame has a high amount of texture related to the current scene. The features extracted from the I-frame are scaled by the distance parameter $\Delta_{t}$, where $0 \leq \Delta_{t} \leq 1$. This allows the I-frame features to have a greater weight if they are closer to the current patch. $\Delta_{t}$ is calculated as : $\Delta_{t} = e^{-0.02(\gamma)}$, where $\gamma$ is the absolute distance between the frame index of input patch and the nearest I-frame. 

The user-defined restoration parameter $d_{t}$ specifies the strength of the restoration and is proportional to the amount of degradation present in the input patch. This value is fed into all stages of the UNet to guide the restoration.  

\begin{table}[]
\caption{\em The impact of varying $\Lambda$ for $C_{opt}$ compared against other critics and VBM4D.}
\label{tab:Results}
\resizebox{\columnwidth}{!}{%
\begin{tabular}{@{}ccccccccc@{}}
\cmidrule(l){2-9}
                                 & \textbf{Lambda} & \textbf{PSNR$_Y$}     & \textbf{PSNR$_{CBCR}$}  & \textbf{SSIM}       & \textbf{VMAF}        & \textbf{LPIPS}      & \textbf{FID}        & \textbf{KID}           \\ \midrule
\multirow{6}{*}{\textbf{C$_{opt}$}} & 0.0             & 26.92                & 33.24                & 0.80                & 46.82                & 0.45                & 0.44                & 3.3e-4                 \\ \cmidrule(l){2-9} 
                                 & 1.0             & 26.07                & 34.82                & 0.79                & 43.67                & 0.45                & 0.44                & 3.5e-04                \\ \cmidrule(l){2-9} 
                                 & 10.0            & 27.98                & 35.06                & 0.79                & {\ul \textit{50.43}} & 0.44                & 0.45                & 3.6e-04                \\ \cmidrule(l){2-9} 
                                 & 100.0           & 28.25                & 36.87                & 0.81                & 48.72                & 0.43                & 0.40                & 3.8e-04                \\ \cmidrule(l){2-9} 
                                 & 1000.0          & 29.11                & 37.75                & 0.83                & 47.28                & {\ul \textit{0.41}} & {\ul \textit{0.36}} & {\ul \textit{3.2e-04}} \\ \cmidrule(l){2-9} 
                                 & 10000.0         & 29.43                & 38.12                & 0.83                & 47.28                & 0.42                & 0.37                & 3.7e-04                \\ \midrule
\textbf{C$_{5,6}$}                & 1000.0          & 29.64                & 38.41                & {\ul \textit{0.83}} & 45.91                & 0.43                & 0.39                & 3.8e-04                \\ \midrule
\textbf{C$_t$}                  & 1000.0          & 29.06                & 37.65                & 0.82                & 49.50                & 0.44                & 0.40                & 3.5e-04                \\ \midrule
\textbf{VBM4D}                   & -               & {\ul \textit{30.86}} & {\ul \textit{39.21}} & 0.82                & 41.09                & 0.43                & 1.2                 & 1.2e-02                \\ \midrule
\textbf{Degraded}                & -               &                      &                      &                     &                      &                     &                     &                        \\ \bottomrule
\end{tabular}}
\end{table}

\begin{figure*}[]
    \centering
    \includegraphics[width=0.95\textwidth]{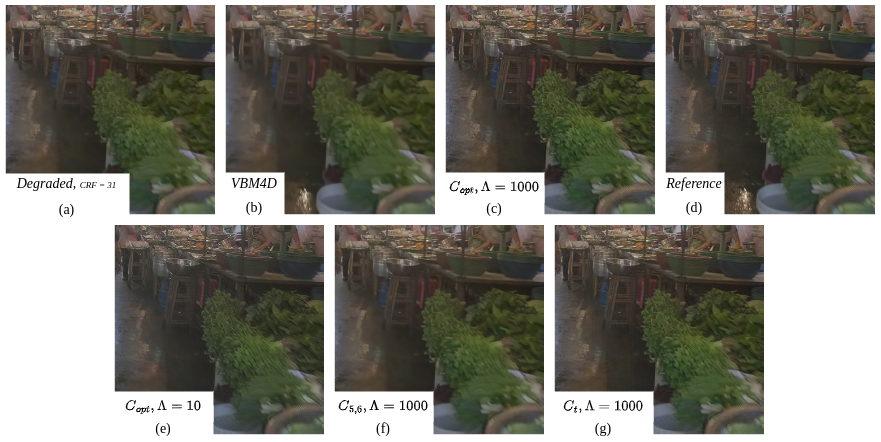}
    \caption{\em Generated patches. VBM4D (b) has the highest PSNR but suffers from oversmoothening. This is a characteristic of optimzing with respect to MSE alone. Our proposed method $C_{opt}$ with $\Lambda$ = 10 yields a high detailed image but the color does not match the original material. $C_{5,6}$ has the highest PSNR of all critic approaches, but still has a loss of detail. $C_{t}$ has higher DFQM scores than $C_{5,6}$, however there is a consistent streaking effect overlaid on degraded parts of the image. $C_{opt}$ with $\Lambda$ = 1000 has the best performing DFQM metrics.}
    \label{fig:results}
\end{figure*}

\section{Training}

A dataset with 2649 degraded/reference pairs was created by degrading clips using the CRF parameter on an H.265 codec such that the resulting PSNR$_{Y}$ of the degraded video was $\leq 38$dB. More details about the dataset can be found in the supplementary material~\cite{suppMaterial}. All training patches were 512x512. 

The degradation parameter $d_{t}$, which controls the strength of the restoration network, was set to be the LPIPS score between the degraded input patch and the reference patch. Our motion magnitude and direction ($m_{t}$) was calculated using the DeepFlow algorithm~\cite{6751282} and the Luma channel of the previous patch, current patch and next patch.

Our generator was initially trained using the squared $l_{2}$ norm loss for 25 epochs. The outputs from this initially trained generator was then used to train the critic network for 10 epochs. This type of training is adapted from the NoGAN approach~\cite{salmona2022deoldify}. The Adam optimizier with learning rate 1e-04 was used in both cases. 
The generator and critic network were then trained adversarially for 50 epochs using Adam optimizers with an initial learning rate of 1e-04 that decayed to 1e-06. Our adversarial training routine follows~\cite{gulrajani2017improved}, with the loss function for both the critic and generator as follows.
\begin{align}
    \mathcal{L}_{critic} & = - \biggl[ C(y_t) - C(G(f_t)) \biggr] + \lambda\Gamma \\ 
    \mathcal{L}_{gen} & = -C(G(f_t)) + \Lambda\bigl(l_2(\hat{f}_t, y_t)\bigr) 
\end{align}
Where $C$, $G$ are the critic and generator network respectively. $\lambda$, $\Gamma$ are the gradient penalty weight and gradient penalty function (as in~\cite{gulrajani2017improved}). $\Lambda$ is a weight that allows the generator to strike a compromise between MSE ($l_2(\cdot, \cdot)$), and perceptual criteria. This is important 
because generative networks like these do have the tendency to synthesise or hallucinate content which {\em looks like} picture content but is not constrained by the input pixel texture.
The MSE weight $\Lambda$ therefore allows the output to be constrained to be related more closely to the input content.

\section{Results}

Table~\ref{tab:Results} shows results of performance of the various systems tested using a unique test set of 501 reference/degraded clips. In addition to the usual performance metrics of VMAF and PSNR, we report Frechet Inception Distance (FID) and Kernel Inception Distance (KID)~\cite{heusel2017gans}. These latter metrics express distance in terms of the output of features from an Inception network. They have recently increased in importance for perceptual evaluation of picture quality as used~\cite{mentzer2022neural}.

In order to select the optimal value of $\Lambda$, the table reports performance for $C_{opt}$ under the effect of  varying $\Lambda$ on a log scale. $\Lambda = 1000$
 is chosen as optimal given that it minimises KID and FID.
Our proposed critic $C_{opt}$ (using perceptual layers) is the best performer with up to 10\% (FID) and 15\% (KID) average improvement  compared to other critics ($C_{5,6}$, $C_{t}$). 

$\Lambda$ is used to weight the effect of the MSE loss function when training the generator. With lower $\Lambda$ the output content has a high level of detail but the color of the generated patches does not match the source material. This effect can be seen in Figure~\ref{fig:results}(e) and is corroborated by the low PSNR$_{Y/CBCR}$ scores. As $\Lambda$ increases the resulting PSNR$_{Y/CBCR}$ also increases as MSE is inversely proportional to PSNR.
The other critic configurations $C_{5,6}$ and $C_{t}$ were also trained with $\Lambda$ = 1000. Patches generated from training with $C_{5,6}$ contain less detail than the other critic configurations, however PSNR$_{Y/CBCR}$ was the best performing among critics. This effect can be seen in Figure~\ref{fig:results}(f).  Training with $C_{t}$ yielded DFQMs that were better performing than $C_{5,6}$, however there are visible streaking effects over areas of high distortion. This can be seen in Figure~\ref{fig:results}(g).
Considering the performance of more traditional statistical approaches, we find our optimal critic to outperform VBM4D in DFQMs. Patches generated from VBM4D suffered from the ``over-smoothing'' effect that is characteristic of MSE optimized algorithms. The visual impact of this is shown in Figure~\ref{fig:results}(b). This validates the need for perceptual criteria.

\section{Discussion \& Conclusion} 
Drawing upon the principles of DFQMs, such as LPIPS, we developed a critic that employs layers from a pre-trained classification network to enhance critic proficiency. Although our model displays lower performance in conventional metrics such as VMAF, PSNR, and SSIM, it generates output that exhibit a close resemblance to the source content and this is reflected in performance increases in LPIPS, FID and KID (Shown in Table~\ref{tab:Results}) as well as visual evidence. 

Our proposed critic $C_{opt}$, outperforms the other critics which indicates that the use of the selected layers from EfficientNetB3 exploits latent information that is invariant to degradation from compression. The superior perceptual performance achieved by our proposed model, as compared to VBM4D, provides evidence of the capactiy of the critic to focus on highly informative perceptual features beyond the conventional distance metrics.

A drawback of pure adversarial training for compression enhancement purposes is the introduction of hallucination artifacts, which can cause the appearance of plausible image features that are not present at all in the source content. To mitigate this, we use a weighted loss function using $\Lambda$ to combine both perceptual critic and MSE. 
Lower values of $\Lambda$ imply more emphasis on perceptual loss and hence result in pictures with greater detail.

In future, we plan to conduct a subjective study to further validate our approach. We also plan to measure the importance of different layers of the critic and the generator. Data, full network architectures and models are available at~\cite{suppMaterial}.
\clearpage
\bibliographystyle{IEEEbib}
\bibliography{cite.bib}

\end{document}